\documentclass[reprint, cha]{revtex4-1}

 \usepackage{graphicx}

\usepackage{gensymb}
\usepackage{amssymb}
\usepackage{amsmath}
\usepackage{textcomp}

\usepackage{xspace}

\usepackage{setspace}

\usepackage{subfigure}
\usepackage{color}
 \usepackage{amsthm}

\begin{document}

\title{Enhancing Fractal Descriptors on Images by Combining Boundary and Interior of Minkowski Dilation}

\author{Marcos William da S. Oliveira}
 	     \email{mw.william@gmail.com}
\affiliation{Institute of Mathematics and Computer Science, University of S\~{a}o Paulo (USP), Avenida Trabalhador s\~{a}o-carlense, 400 13566-590 S\~{a}o Carlos, S\~{a}o Paulo, Brazil} 

\author{Dalcimar Casanova}
 	     \email{dalcimar@gmail.com}
\affiliation{Scientific Computing Group, S\~ao Carlos Institute of Physics, University of S\~{a}o Paulo (USP),  cx 369 13560-970 S\~{a}o Carlos, S\~{a}o Paulo, Brazil - www.scg.ifsc.usp.br} 

\author{Jo\~ao B. Florindo}
 	     \email{florindo@ursa.ifsc.usp.br}
\affiliation{Scientific Computing Group, S\~ao Carlos Institute of Physics, University of S\~{a}o Paulo (USP),  cx 369 13560-970 S\~{a}o Carlos, S\~{a}o Paulo, Brazil - www.scg.ifsc.usp.br} 

\author{Odemir M. Bruno}
              \email{bruno@ifsc.usp.br}
\affiliation{Scientific Computing Group, S\~ao Carlos Institute of Physics, University of S\~{a}o Paulo (USP),  cx 369 13560-970 S\~{a}o Carlos, S\~{a}o Paulo, Brazil - www.scg.ifsc.usp.br}

\date{\today}

\begin{abstract}
This work proposes to obtain novel fractal descriptors from gray-level texture images by combining information from interior and boundary measures of the Minkowski dilation applied to the texture surface. At first, the image is converted into a surface where the height of each point is the gray intensity of the respective pixel in that position in the image. Thus, this surface is morphologically dilated by spheres. The radius of such spheres is ranged within an interval and the volume and the external area of the dilated structure are computed for each radius. The final descriptors are given by such measures concatenated and subject to a canonical transform to reduce the dimensionality. The proposal is an enhancement to the classical Bouligand-Minkowski fractal descriptors, where only the volume (interior) information is considered. As different structures may have the same volume, but not the same area, the proposal yields to more rich descriptors as confirmed by results on the classification of benchmark databases.
\end{abstract}


\maketitle

\section{Introduction}

In the last decades, Fractal Geometry has demonstrated to be a worthy tool to develop robust and precise methods of image analysis \cite{SMS10,HWZ08,CDHLAB03,W08,TWZ07,LC10}. Such methods have been successfully applied to a number of problems comprising the analysis of digital images in several areas, such as Physics \cite{POPMBD97,BOAM10}, Medicine \cite{KSASKC12,GQPJ11}, Engineering \cite{ACNS09,DOSSS09}, etc.

Among such fractal-based approaches, methods like multifractals \cite{XJF09}, multiscale fractal dimension \cite{MCSM02}, local fractal dimension \cite{K99}, etc. have outperformed other classical and state-of-the-art image analysis methods in many situations.

More recently, a novel fractal-based imaging method named fractal descriptors has been proposed in \cite{BCB09,FB11}. Roughly speaking, this approach extracts features of the image by computing the fractal dimension at different scales of observation and taking all these values over a predefined range. Although this approach has demonstrated to be a promissing solution for image analysis problems, it was defined and studied only for a limited number of well-known techniques to estimate the fractal dimension. Among the studied possibilities, Bouligand-Minkowski has provided remarkable results both for the analysis of natural and synthetic images \cite{BCB09}. These descriptors are obtained from the interior measures (volumes) of the object of interest dilated by spheres with a predefined range of radius values.

Despite the great results achieved, Bouligand-Minkowski takes into account only the interior (volume) of the dilated structure. It is well-known from Geometry that the boundary of a structure encloses information as rich as the interior and, for instance, in a three-dimensional space, two objects with the same volume may have different areas. In this way, this work proposes to enhance the Bouligand-Minkowski descriptors by including information from the boundary, that is, the area in a three-dimensional space. The combination is accomplished by means of a simple concatenation of measures, followed by a dimensionality reduction through the cannonical analysis \cite{CorreaSC10}. The performance of the proposal is assessed over databases of texture images and the results are compared to other classical and state-of-the-art methods in the literature.

\section{Fractal Geometry}

A number of works characterizing and analyzing images using fractal geometry have been reported in the literature \cite{SMS10,HWZ08,CDHLAB03,W08,TWZ07,LC10}. Generally, these works model objects and scenarios from the real world as an approximation of mathematical or statistical fractals and extract fractal properties of the element. The most used of these properties is the fractal dimension.

The formal definition of the fractal dimension, also called Hausdorff-Besicovitch dimension, is obtained from the Hausdorff measure. Let $X$ be a geometrical set of points in an $N$-dimensional topological space. Its Hausdorff measure $H^{s}_{\delta}$ is calculated by
\begin{equation}
	H^{s}_{\delta}(X) = \inf{\sum_{i=1}^{\infty}{|U_{i}|^{s}}},
\end{equation}
where $U_{i}$ is a $\delta$-cover of $X$; that is, there exists a countable collection of sets $\{U_i\}$, with $|U_i| \leq \delta$, such that $X \subset \cup_{i=1}^{\infty}U_i$ and $|U_i|$ denotes the diameter of $U_i$, that is, the maximum possible distance between two any elements of $U_i$:
\begin{equation}
	|U_i| = \sup\{\|x-y\|:x,y \in U_i\}.
\end{equation}

As the parameter $\delta$ is a superior limit for the diameter of the balls $U_i$ covering the fractal object, it can be considered a scale metric and should be removed from the dimension definition once this is scale-independent. In this way a limit to $0$ is applied over $\delta$ giving rise to the measure $H^s$
\begin{equation}
	H^s(X) = \lim_{\delta \rightarrow 0}H^s_{\delta}(X).
\end{equation}

As it can be demonstrated in Measure Theory, $H^s(X)$ has a particular behavior that arises for any set of points $X$, that is, the value of $H^s$ is always $\infty$ for any $s < D$ and $0$ for any $s > D$, where $D$ is a non-negative real value. The point of discontinuity $D$ is the Hausdorff-Besicovitch fractal dimension of $X$
\begin{equation}
	D(X) = \inf \left\{ s:H^{s}(X)=0 \right\} = \sup \left\{ s:H^{s}(X)=\infty \right\}.
\end{equation}

Although the above definition is the most exact and generalist method to calculate the dimension, by using an infinitesimal covering, it is necessary to know the analytical expression of the object being measured. However this is not possible when the real-world element approximated by a fractal is represented in a discrete and finite space as the digital images discussed in this work. To address these situations, several approximation methods have been proposed in the literature \cite{F86,FB11b}. Such methods aim to compute a measure of self-similarity and complexity of the object by generalizing the definition of the Euclidean dimension. Thus the object is measured by a \textit{rule unit} with the same topological dimension of the object. The length $r$ of this unit is ranged along an interval to compute the number of units $N(r)$ necessary to cover the object. In this context, the dimension is given by
\begin{equation}\label{eq:FD}
	D = -\lim_{r \rightarrow 0}\frac{\log(N(r))}{\log(r)}.
\end{equation}

\subsection{Bouligand-Minkowski}

One of the most commonly used techniques to estimate the dimension based on Equation \ref{eq:FD} is the Bouligand-Minkowski \cite{F86}. Similarly to the Hausdorff-Besicovitch definition, it is also derived from a measure, in this case, the Bouligand-Minkowski measure $M(X,S,\tau)$ of a set $X \in \mathbb{R}^n$
\begin{equation}
	M(X,S,\tau) = \lim_{r \rightarrow 0}\frac{V(\partial X \oplus S_r)}{r^{n-\tau}},
\end{equation}
where $-\infty < \tau < +\infty$ is a parameter and $V(\partial X \oplus S_r)$ is the volume of the edge of $X$ ($\partial X$) morphologically dilated by a structuring element $S$, symmetrical with respect to the origin and with radius $r$.

The Bouligand-Minkowski fractal dimension $D_{BM}$ is given by
\begin{equation}
	D_{BM}(X,S) = \inf \left\{ \tau : M(X,S,\tau) = 0 \right\}.
\end{equation}

In practice, the dimension is computed by a neighborhood strategy. Each point of the object $X$ is replaced by a structuring element $S_{\epsilon}$, with radius $\epsilon$, and the number of points within the union of such elements is used to estimate the volume $V$. Thus the dimension is provided by
\begin{equation}
	D_{BM}(X) = \lim_{\epsilon \rightarrow 0}\left( n-\frac{log V(X \oplus S_{\epsilon})}{log \epsilon} \right).
\end{equation}

For an object represented in a digital image, an efficient and precise method to compute the Bouligand-Minkowski dimension is the Euclidean Distance Transform \cite{FCTB08}, by Saito's Algorithm \cite{SaitoT94}. Starting from a texture image $I:\mathbb{R}^2 \rightarrow \mathbb{R}$, it is mapped into a 3D structure (surface) $X$, such that each point with coordinate $(x,y)$ and pixel intensity $z$ is converted into the point with coordinate $(x,y,z)$ in $X$. To simplify the idea, the object $X$ is supposed to be in $\mathbb{R}^3$. The distance transform $DT_X$ is given by
\begin{equation}
	DT_X(i,j)=\min\{d((i,j,k),(i',j',k')): (i',j',k')\in X\},
\end{equation}
for all $(i,j,k)\in\mathbb{R}^3$, where $d(p,q)$ is the Euclidean Distance between $p$ and $q$.

\begin{figure}[!htpb]
\centering
\includegraphics[width=\columnwidth]{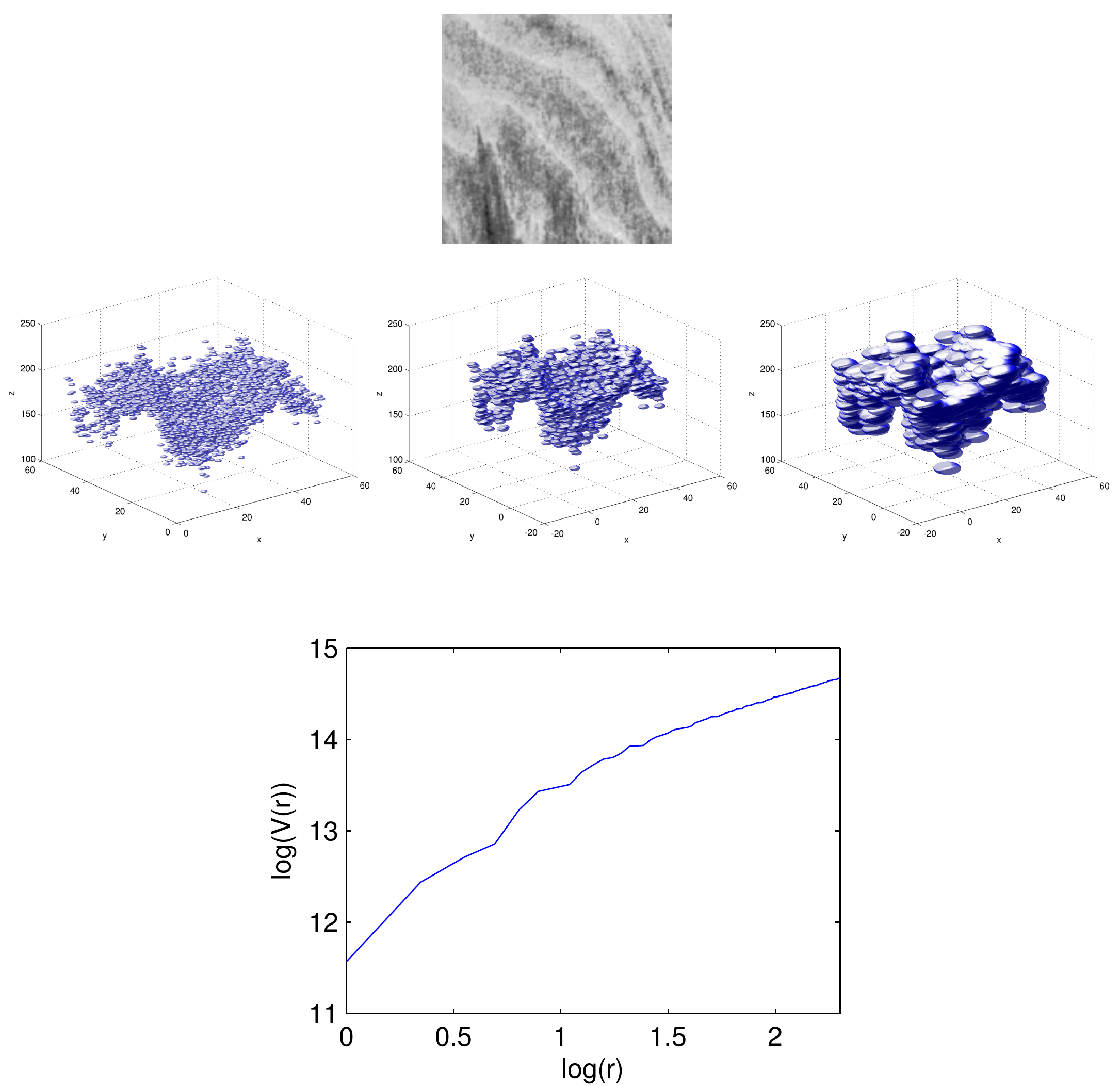}
\caption{Bouligand-Minkowski fractal dimension for three-dimensional objects. A texture image is mapped onto a surface and handled as a three-dimensional structure \cite{BCB09}. The $\log(r) \times \log(V(r))$ curve can be used to estimate the fractal dimention of the object.}
\label{fig:mink}
\end{figure}

The set of possible distances $R$ (Euclidean) is given by
\begin{equation}
	R = \{r : r = \sqrt{i^2+j^2+k^2}; i,j,k \in \mathbb{N}\}.
\end{equation}
In the following, these values of $r$ are sorted increasingly
\begin{equation}
	R = \{r_0,r_1,r_2,r_3,r_4,...,r_{max}\} = \{0,1,\sqrt{2},\sqrt{3},2\sqrt{2},...,r_{max}\}.
\end{equation}
Thus the dilation volume $V(r)$ is computed by
\begin{equation}
	V(r) = \#\{(i,j,k) : DT_X(i,j,k)=r\}
\end{equation}
and the dimension is given by
\begin{equation}
	D_{BM} = 3 - \lim_{r \rightarrow 0}\frac{\log(V(r))}{\log(r)}.
\end{equation}
Numerically, this limit uses to be given by the slope of a straight line fit to the curve $\log(r) \times \log(V(r))$. Figure \ref{fig:mink} illustrates the dilation process.

\subsection{Fractal Descriptors}

Equation \ref{eq:FD} can be generalized by replacing $N(r)$ by any self-similarity measure $\mathfrak{M}(\epsilon)$, where $\epsilon$ is a scale parameter.

This generalization results in the proposal and study of several methods to approximate the dimension \cite{F86,FB11b}, each one providing results more or less close to the theoretical value, depending on the specific application. However, the outcome still is only a single real value to describe all the complexity of an object. Moreover, when this object is not a mathematical fractal, its dimension is highly scale-dependent and a global dimension may be of little or no usefulness.

To address these points and make possible a more complete fractal-based analysis of real-world structures, the fractal descriptors were proposed in \cite{BCB09,FB11}. Basically, instead of computing only the fractal dimension, the fractal descriptors are composed by all the values of dimension at each scale along a range of observation. Considering that $\epsilon$ is a scale parameter, such set of descriptors $u$ can be obtained from the self-similarity curve:
\begin{equation}
	u : \epsilon \rightarrow \mathfrak{M}(\epsilon).
\end{equation}
Using the Bouligand-Minkowski method described previously, in thres dimensions, these descriptors are given by the logarithm of the dilation volumes
\begin{equation}
	u = [\log(V(r_0)),\log(V(r_1)),...,\log(V(r_{max})))].
	\label{eqt:fracDesc}
\end{equation}

In an image analysis task, the descriptors $u$ can be used directly \cite{BCB09} or after some type of transform \cite{FCB11} as well as they can be extracted from the entire image \cite{FB11} or from disjoint regions \cite{FB13}.

\section{Boundary Measure \textit{versus} Interior Measure}

On the above discussion, the Equation \ref{eqt:fracDesc} and the previous works \cite{BCB09,FB11} talk about the use of interior information (volume in three dimensional space) to obtain the fractal descriptors. This work proposes to analise the difference between the information comprised within the boundary and the interior of a region. In two dimensions, this analysis consists of studying measures of perimeter and area of a flat shape. In the three-dimensional case, the discussion concerns area measures of surfaces and measures of volumes.

The main idea depicted here is that two different objects, with a dilatated radius $r$, can have the same interior measure, but different boundary measures.

\begin{figure}[!htpb]
\centering
\includegraphics[width=\columnwidth]{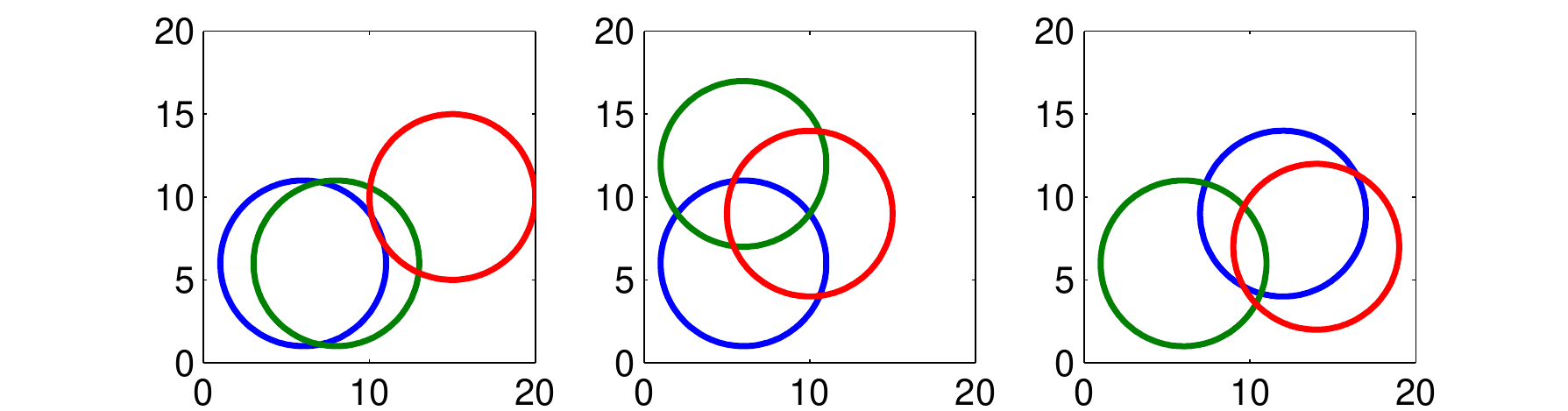}
\caption{Different arrangements of three circles with the same radius. The center points, areas and perimeter measures are explained on Table \ref{tab:AreavsPerimeter}.}
\label{fig:circles}
\end{figure}

To support this theory, the dilation of circles is analysed in two dimensions. Figure \ref{fig:circles} illustrates three different circle arrangements, with three different center points for each case and the same radius. It is an exemple that the interior areas of three circles, including the intersections, are the same but the the boundary measures of the regions, i.e. the perimeters, are different.

In Image Processing, the perimeter of a flat region $\Theta$ can be estimated by
\begin{equation} \label{eq:Perimeter}
	P = n_e+n_o\sqrt{2},
\end{equation}
where $n_e$ and $n_o$ are the number of even and odd codes, respectively, in chain-code representation \cite{CC00,C96}. Equation \ref{eq:Perimeter} simply performs a count of the number of pixels on the boundary of a region and estimates the arc length of its contour.

In the same way, the area measure is estimated by the count of pixels that represent the object  $\Theta$. Therefore, the area is obtained by

\begin{equation} \label{eq:Area}
	A = \sum_{p\in\Theta}1,
\end{equation}
where $p$ denotes any pixel on the image.

After applying Equations \ref{eq:Area} and \ref{eq:Perimeter} for the three arrangements on Figure \ref{fig:circles}, similar area measures and different perimeters are obtained. Table \ref{tab:AreavsPerimeter} shows these results to the arrangements of the Figure \ref{fig:circles}.

\begin{table*}[ht]	
	\centering
		\begin{tabular}{c|c|c|c}
			\hline						
			 Arrangements & Center Points & Area & Perimeter  \\
			\hline
            Left   & (6,6); (8,6); (15,10)  & 152 & 49.799 \\
            Center & (6,6); (6,12); (10,19) & 152 & 44.627 \\
            Right  & (6,6); (12,9); (14,7)  & 152 & 46.627 \\
			\hline
		\end{tabular}
	\caption{Center points, interior areas and perimeters of three arrangements of circles on Figure \ref{fig:circles}.}
	\label{tab:AreavsPerimeter}
\end{table*}

These results support our theory, we can verify that the two different images (or pixels arrangement) can have the same dilated areas (to a given dilatated radius $r$), but different perimeters. In this situation only the perimeter is useful to discriminate the images. In this way we can use the bondary measure as an complementary feature to compose our fractal descriptor. 

For three-dimensional objects, the surface area measure and volume measure are estimated in a similar way to that of Equation \ref{eq:Perimeter} and \ref{eq:Area}. Furthermore, the behavior of sphere dilation is also similar to that of circle dilation.


\section{Applications on Images}

As showed by the above discussion, the area and perimeter of a dilated object express complementary information. In this way the most natural approach is to use both features together. The synergy achieved by this approach can bring a better discrimination power to the fractal descriptors. To do that, a simple concatenation of both feature vectors is made in order to obtain a more rich fractal descriptors.

In order to verify our theory an application of texture analysis is carried out using the novel proposed fractal descriptor. Such analisys is performed over a supervised classification task, though the proposed method can be also used to perform a CBIR (Content-based image retrieval), segmentation or other kinds of image analysis.

For this application, the both signatures (area and volume) are computed for each image, concatened and the supervised classification is carried out by applying a Canonical Analysis \cite{CorreaSC10} followed by a Linear Discriminant Analysis (LDA) (also called Fisher linear discriminant) \cite{Fukunaga90}. The 10-fold cross-validation scheme is used in all experiments and over benchmark databases. The canonical analysis is, basically, a geometric transformation of the feature space in order to generate new uncorrellated features based on linear combinations. The idea of this method is to find a new projection of the data where the class separation is maximized. From $p$ original features, $p$-cannonical variables can be obtained. However, a reduction in the number of variables to be evaluated is usually desired. Therefore, a LDA supervised classification is accomplished by using the most significant $p$ cannonical variables.

\subsection{Texture}

Three texture sets are used in the experiments: Brodatz, Vistex and Outex.

\begin{enumerate}
	\item Brodatz texture database is derived from the Brodatz Album \cite{Brodatz66} and has become the standard for evaluating texture algorithms, with hundreds of studies having been applied to this set of images. This database is composed by $1776$ texture samples grouped into $111$ classes. Each image is $128 \times 128$ pixels with 256 gray-levels.

	\item The Vision texture database (VisTex) \cite{Vistex09} is maintained by the Vision and Modelling group at the MIT Media Lab. The full database contains images representative of real-world textures under practical conditions (lighting, perspective, etc). In this work, we use the $54$ original Vistex images with resolution $512\times512$. Each image was split into $16$ non-overlapping sub-images with dimension $128\times128$. These images are avaliable on de Vistex site as the test suite Contrib\_TC\_00006 \cite{OjalaMPVKH02}.

	\item The suite Outex\_TC\_00013 is provided by the Outex texture database \cite{OjalaMPVKH02}. This database includes a collection of natural scenes acquired under strictly controlled conditions. The test suite provides a meaningful framework for the empirical evaluation of a candidate texture analysis algorithm. A database of $1360$ color texture images ($128\times128$) was constructed by splitting each one of the $68$ original texture image ($746\times538$) into $20$ non-overlapping sub-images.
\end{enumerate}

\begin{figure}[!htpb]
	\centering
		\includegraphics[width=\columnwidth]{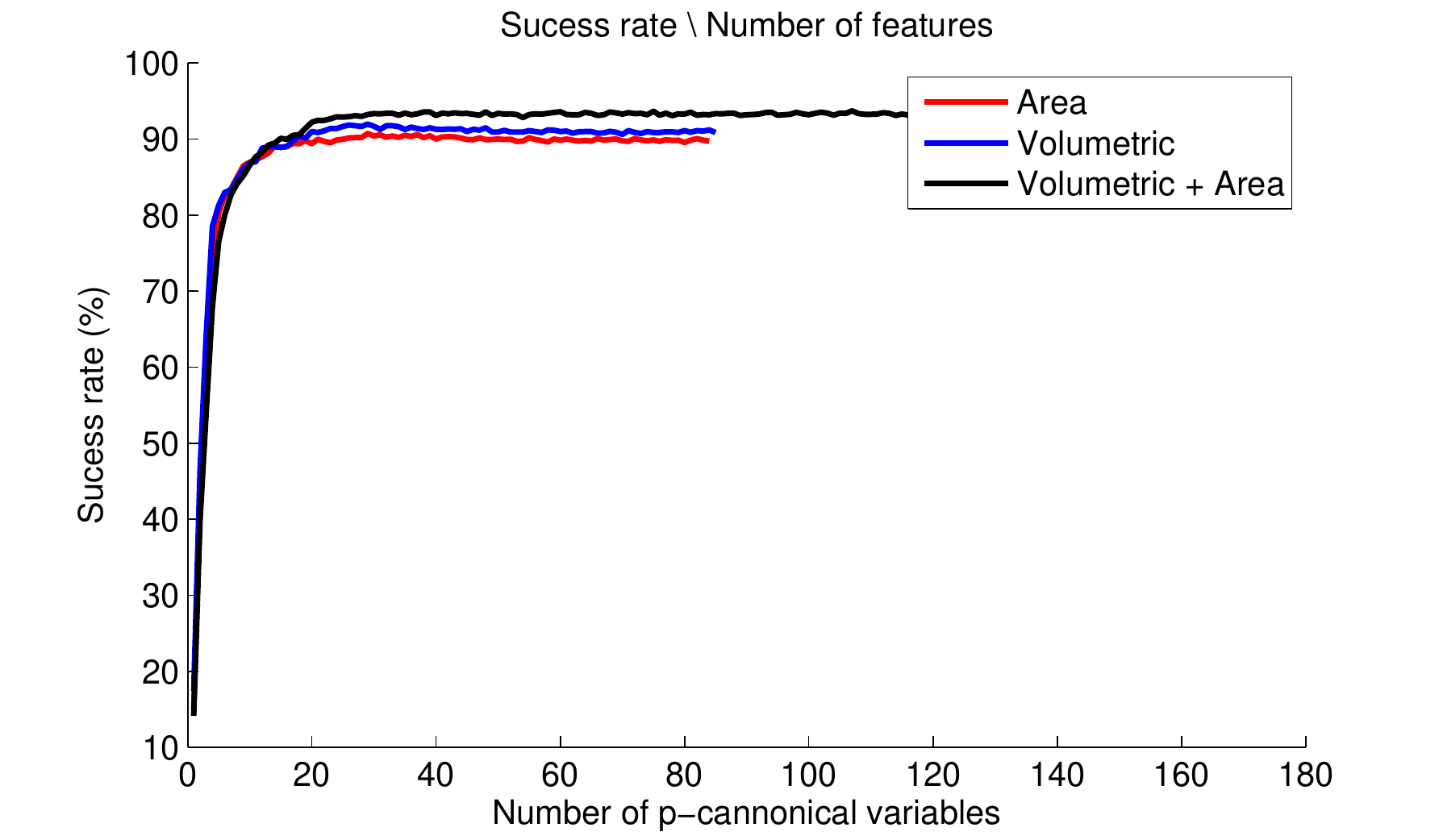}
	\caption{Accuracy \emph{versus} number of $p$-cannonical variables. The optimal $p$-cannonical variables of Vistex datasets is nearly 30-40 features.}
	\label{fig:rateXfeatures}
\end{figure}

After the concatenation and Canonical Analysis the success rate may vary depending on the number of the $p$-cannonical variables used in LDA classifier. Figure \ref{fig:rateXfeatures} illustrates the behavior of the success rates when the number of $p$-cannonical variables is ranged. We observe that, for Vistex data set, the rate increases at a first moment, achieves an optimal rate and thus stabilizes with a very small decrease when we consider more descriptors. This behavior was expected since the high dimensionality of the feature vectors damage the efficiency of the classifier. For all considered fractal descriptors this same behavior can be observed.

We also can verify that the volumetric information is a little bit more effective than area counterpart. However, we see that the synergy between such two kinds of infomation yields a improvement on the perfomace of the texture analysis.

Based on the behavior observed in Figure \ref{fig:rateXfeatures}, we setup an experiment with a total of 40 $p$-cannonical variables. Table \ref{tab:Comparison} shows the achieved results. For all data sets, the best sucess rate is provided by the combination of volumetric and surface features, the average performance improvement is close to 2\%. This result corroborates our theory that the area can be useful as a complementary information in the studied fractal descriptor. It is important to emphasize that combining different features does not increase the number of descriptors used in the classifier.

\begin{table*}[ht]	
	\centering
		\begin{tabular}{c|c|c|c}
			\hline						
			 & \multicolumn{3}{c}{Success rate (\%) and stardart deviation (\%)}\\
			\cline{2-4}			
			Methods & Brodatz & VisTex & Outex \\
			\hline
				Volumetric &  				88.15(0.26) & 91.28(0.67) & 80.66(0.43) \\
				Area & 								87.93(0.29) & 89.98(0.52) & 80.91(0.37) \\
				Volumetric + Area &  	89.29(0.21) & 93.09(0.31) & 82.57(0.33) \\
			\hline
		\end{tabular}
	\caption{Results of fractal descriptors on texture databases using 40 $p$-cannonical variables. Note that the synergy between the volumetric and area features can improve the final results.}
	\label{tab:Comparison}
\end{table*}

\section{Conclusion}

This work proposed a new way of computing fractal descriptors from gray-level fractal descriptors. The method combines information from the interior (volume) and boundary (area) of a surface representation of the image dilated by spheres with variable radii (Minkowski dilation).

The results of applying the proposal to the classification of benchmark data sets showed that by using a reasonable number of descriptors, the volume achieved higher correctness rates than the area used in the Bouligand-Minkowski approach. However the performance is enhanced when combining area and volume information. 

In fact, the area enriched the descriptors in that it provides a different and complementary viewpoint of the texture. Thus different images can have the same volume for a specific dilation, but its area may be different, contributing for a more robust discrimination.

\section*{Acknowledgments}
O. M. Bruno gratefully acknowledges the financial support of CNPq (National Council for Scientific and Technological Development, Brazil) (Grant \#308449/2010-0 and Grant \#484312/2013-8) and FAPESP (The State of S\~ao Paulo Research Foundation) (Grant \# 11/01523-1).
M. W. S. Oliveira is thankful to CAPES (Coordination for the Improvement of Higher Education Personnel) for financial support.
J. B. Florindo gratefully acknowledges the financial support of FAPESP Proc. 2013/22205-3 and 2012/19143-3.
D. Casanova is grateful to FAPESP Proc. 13/14984-2 for financial support.


\end{document}